\documentclass{article}
\usepackage[margin=1in]{geometry}

\usepackage[numbers]{natbib}

\usepackage{authblk}

\usepackage{amsmath}

\usepackage{moreverb,url}

\usepackage[colorlinks,bookmarksopen,bookmarksnumbered,citecolor=red,urlcolor=red]{hyperref}

\usepackage{multirow}
\usepackage{subcaption}
\usepackage{booktabs}
\usepackage{tabularx}

\usepackage{graphicx}

\title{Integrating Misclassified EHR Outcomes with Validated Outcomes from a Non-probability Sample}

\author[1]{Jenny Shen\thanks{jenshen@pennmedicine.upenn.ed}}
\author[1]{Dane Isenberg}
\author[1]{Kristin A. Linn}
\author[2]{Rebecca A. Hubbard}

\affil[1]{Department of Biostatistics, Epidemiology, and Informatics, Perelman School of Medicine, University of Pennsylvania, PA, USA}

\affil[2]{Department of Biostatistics, Brown University School of Public Health, RI, USA}

\usepackage{bm}

\begin{document}

\maketitle

\begin{abstract}
Although increasingly used for research, electronic health records (EHR) often lack gold-standard assessment of key data elements. Linking EHRs to other data sources with higher-quality measurements can improve statistical inference, but such analyses must account for selection bias if the linked data source arises from a non-probability sample. We propose a set of novel estimators targeting the average treatment effect (ATE) that combine information from binary outcomes measured with error in a large, population-representative EHR database with gold-standard outcomes obtained from a smaller validation sample subject to selection bias. We evaluate our approach in extensive simulations and an analysis of data from the Adult Changes in Thought (ACT) study, a longitudinal study of incident dementia in a cohort of Kaiser Permanente Washington members with linked EHR data. For a subset of deceased ACT participants who consented to brain autopsy prior to death, gold-standard measures of Alzheimer's disease neuropathology are available. Our proposed estimators reduced bias and improved efficiency for the ATE, facilitating valid inference with EHR data when key data elements are ascertained with error.\\

\setlength{\parindent}{0pt}\textbf{Keywords}: Electronic health records, Data integration, Measurement error, Selection bias, Alzheimer's disease

\end{abstract}


\section{Introduction}
\label{s:intro}

Although electronic health records (EHRs) were not originally collected for research purposes, these health care-derived data resulting from administration and delivery of clinical care have been adopted and used increasingly in clinical and epidemiological studies.\cite{callahan2020research, farmer2018promises} The passage of the Health Information and Technology for Economic and Clinical Health (HITECH) Act of 2009 \cite{adler2014more} facilitated greater access to the abundance and wealth of information collected in these records. At a comparatively low cost, biomedical investigators can now query information on millions of patients through EHRs, link patient information to other biomedical data such as genomics, and attempt to leverage this information for a variety of research purposes.\cite{agniel2018biases} EHRs have been used for studying and estimating prevalence, risk factors, or progression of disease; informing prescription choices for medication; and guiding determinations for environmental hazards or health policy reforms.\cite{farmer2018promises,verheij2018possible} Furthermore, EHRs offer the benefit of studying populations over a longer term than may be possible in a clinical trial or to study individuals who may be underrepresented within randomized clinical trials. However, given that EHRs were not purposefully collected for research and are prone to irregular sampling, missingness, unmeasured confounding, and other data quality issues, EHR-based analyses must take care to address the complexities inherent in these records to avoid drawing biased or misleading conclusions.\cite{capurro2014availability, agniel2018biases, farmer2018promises,callahan2020research, hubbard2021effective}  

In this work, we consider the context in which an EHR-derived sample may be considered a probability sample of the underlying patient population treated in the healthcare system, and it is supplemented with data from a non-probability sample that includes higher quality outcome assessment for a subset of individuals in the EHR database. Utilizing the non-probability sample may help to address measurement error in EHR-derived outcomes but may introduce selection bias. Our approach is motivated by the Adult Changes in Thought (ACT) study, a longitudinal study of incident dementia conducted among individuals randomly selected from the Kaiser Permanente (KP) Washington \cite{j2012adult} health system. We assume individuals within ACT make up a simple random sample of KP Washington members.\cite{kukull2002dementia} ACT has been used to study Alzheimer’s Disease (AD) which is clinically diagnosed through a combination of cognitive assessment, biomarker measurement, and brain imaging. However, clinical AD diagnoses represent silver-standard assessment of AD and may not correspond to true underlying disease status. Gold-standard diagnosis of AD can only be ascertained via neuropathologic assessment obtained from post-mortem brain autopsy. A subset of individuals in the ACT study consented to autopsy, thereby forming a validation sample containing gold-standard outcomes. Leveraging these gold-standard outcomes in the validation sample could lead to improved inference for the larger EHR sample. However, as Haneuse et al.\cite{haneuse2009adjustment} highlighted, the autopsy cohort is subject to selection bias, so analyses involving this cohort must account for the potential non-representativeness of the subset of ACT participants with available autopsy data. Thus, this scenario warrants an approach for integrating data that accounts for both measurement error in the larger EHR sample and sample selection bias in the non-probability sub-sample.

Three approaches commonly used for data integration are mass imputation, propensity score-based weighting, and calibration weighting. data set for imputing missing values to all units in the probability sample (see, for instance, Kim and Rao \cite{kim2012combining}). Methods such as propensity score-based weighting and calibration weighting are based on causal inference approaches and may be leveraged for estimating parameters such as the average treatment effect (ATE). The ATE is often a parameter of interest in observational studies and was of interest in our study for estimating the effect of hypertension on developing AD. With propensity score-based weighting methods, selection bias is addressed by modeling and estimating the probability of selection into a non-probability sample, i.e., propensity score for selection.\cite{elliott2017inference,chen2020doubly} Calibration weighting produces a weighted distribution in the non-probability sample that is similar to a target population by forcing the auxiliary variables of the probability and non-probability samples to have the same moments or empirical distribution.\cite{disogra2011calibrating} Estimators that combine outcomes from a small probability sample and large non-probability sample also have been shown to yield estimates with greater accuracy and smaller mean squared error (MSE) relative to using only gold-standard outcomes from the probability sample (see, for example, Elliott and Haviland \cite{elliott2007use} and the ``blended calibration" approach from Disogra et al.\cite{disogra2012using}). A recent review article \cite{shi2023data} also noted reduced bias or improved efficiency from implementing methods that integrate observational data with trial data \cite{yang2020improved, kallus2018removing, gui2024combining,athey2020combining} or validation data that forms a random sample of the target population.\cite{yang2019combining}

Measurement error and sample selection bias are two issues that feature prominently in EHR-based analyses and must be addressed appropriately when using EHR data. Measurement error in EHR data may arise in the covariates, outcomes or both, but the majority of causal inference methods have focused on measurement error for covariates, whether for baseline covariates \cite{mccaffrey2013inverse, rudolph2018using} or primary exposure.\cite{babanezhad2010comparison, braun2016using} Shu and Yi \cite{shu2019causal} proposed a method that accounts for measurement error in the outcome, but in the case where validation data are available from a simple random sample (SRS). Performance of this approach has not been evaluated in the setting where the validation data arise from a non-probability sample. Other approaches have sought to address measurement error and selection bias in EHR simultaneously but have focused on other estimands or selection bias for the EHR sample as a whole.\cite{jurek2013adjusting, beesley2022statistical, zeng2023causal} Currently, we are unaware of any approaches to estimation of the ATE that simultaneously leverage gold-standard binary outcomes from a validation sample  while accounting for both mismeasurement in the EHR-derived outcome and the potential for selection bias into the validation sample.

In this paper, we propose a method for inference for the average treatment effect when integrating a large probability sample subject to outcome measurement error with a small non-probability sample that contains gold-standard outcomes. We propose estimators for integrating data from these samples in a way that minimizes bias and leverages information from both samples to improve statistical inference. We compare the performance of our proposed estimators to relevant existing estimators using simulation studies across various data generating mechanisms for the larger probability sample and validation sample. In Section \ref{s:realdata}, we apply the proposed estimators to study the effect of hypertension on development of AD neuropathology using data from the ACT study.

\section{Methodology}
\label{s:methods}
\subsection{Estimation of the ATE}
We assume the target of inference is the ATE and focus on inverse-probability weighted (IPW) estimators for the ATE. Our focus on IPW estimators is motivated by their ease of implementation and interpretability.\cite{shu2019causal} Let $T$ denote an observed binary treatment or exposure variable and $X$ denote pre-treatment covariates. Let $Y_1$ and $Y_0$ represent the potential outcomes that would have been observed if a subject had experienced treatment $T=1$ or $T=0$, respectively. The ATE is defined as $\tau = E(Y_1 - Y_0)$. This causal effect can be identified assuming the standard set of causal inference assumptions of ignorability, positivity, and consistency.\cite{lunceford2004stratification} These assumptions are as follows:
\begin{enumerate}
    \item Under the ignorability assumption, the potential outcomes are independent of treatment assignment, possibly conditional on a set of variables $X$, $(Y_1, Y_0) \perp T| X$.
    \item Assuming positivity, $0 < P(T=1 | X) < 1$ for all $X$.
    \item Assuming consistency of treatment, $Y = TY_1 + (1-T)Y_0$. We define $e = P(T=1 | X)$ to be the probability of receiving treatment $T=1$ given $X$. 
\end{enumerate}

\noindent Given these assumptions, it can be shown that $\tau = E(Y_1 - Y_0) = E_X[E(Y | T=1,X=x) - E(Y | T=0,X=x)]$. 

Without accounting for measurement error in the outcome and selection bias in the validation sample, the IPW estimate of the ATE \cite{rosenbaum1998} is $\hat{\tau} = \frac{1}{n}\sum_{i=1}^n \frac{T_i Y_i}{\hat{e_i}} - \frac{1}{n}\sum_{i=1}^n \frac{(1-T_i) Y_i}{(1-\hat{e_i})}$, where $\hat{e_i}$ is an estimate of $P(T_{i}=1 | X_{i})$. With misclassified outcomes, denoted by $Y^*$, a naive estimator of the ATE would be $\hat{\tau}^* = \frac{1}{n}\sum_{i=1}^n \frac{T_i Y_i^*}{\hat{e_i}} - \frac{1}{n}\sum_{i=1}^n \frac{(1-T_i) Y_i^*}{(1-\hat{e_i})}$. When misclassification in the outcome is present or suspected and validation data containing true outcomes are available, a number of estimators may be considered -- including our newly proposed estimators -- which are detailed in the next section. Because we assume no measurement error in $T$ or $X$, the observed data can be used to posit a model for treatment propensity $e = P(T=1 | X)$ and obtain an estimate, $\hat{e}(X)$ for any $X$. To simplify notation, we have dropped the dependence on $X$ and used $e_i$ and $\hat{e}_i$ to denote subject $i$'s true and estimated propensity of treatment, respectively.   

Let $\bm{Y}$ and $\bm{Y}^*$ denote vectors containing true and error-prone outcomes, respectively, for the entire sample, noting that elements of $\bm{Y}$ for individuals not in the validation data will be missing. We use subscript $\mathcal{V}$ or $\mathcal{V}^C$ to denote the set of vector indices corresponding to subjects in the validation sample and the complement of the validation sample, respectively. In other words, $\mathcal{V} = \{i: Y_i \text{ observed}\}$ and $\mathcal{V}^{C} = \{i: Y_i \text{ not observed}\}$. Let $V_i = I\{i \in \mathcal{V}\}$ be an indicator that takes value 1 if individual $i$ is included in $\mathcal{V}$ and 0 otherwise. We will denote all estimators using modifications of the general form $\tau(\cdot, \cdot)$, where the first and second arguments will denote the subset of $\bm{Y}^*$ and $\bm{Y}$ used by the estimator, respectively. Furthermore, estimators that do not take into account sample selection propensities for estimating the ATE will be denoted by $\hat{\tau}(\cdot, \cdot)$, while estimators that do incorporate sample selection propensities will be denoted by ${\hat{\tau}^S}(\cdot, \cdot)$. For example, $\hat{\tau}(Y_\mathcal{V}, Y^{*}_{\mathcal{V}^C})$ denotes an estimator that: 1) uses all true outcomes from the validation sample; 2) uses mismeasured outcomes from only the individuals not in the validation sample; and 3) does not incorporate sample selection propensities. Estimators without a first or second argument will denote estimators that do not incorporate any of the gold- or silver-standard outcomes, respectively. For example, $\hat{\tau}^S(Y_\mathcal{V}, \cdot)$ will denote an estimator that incorporates selection propensities in estimation of the ATE but only uses the gold-standard outcomes from the validation data. 


\subsection{Handling Outcome Misclassification Using a Validation Sample}

Shu and Yi \cite{shu2019causal} proposed IPW estimators for the ATE in the presence of mismeasured outcomes assuming that the validation sample is a simple random sample (SRS). Misclassification probabilities can be used to correct for outcome mismeasurement but are often unknown in practice. Validation samples provide one useful setting to estimate misclassification probabilities. Consider a validation sample of size $n_{\mathcal{V}}$ containing $X, T, Y, \text{and } Y^*$ for all $n_{\mathcal{V}}$ individuals. Let $p_{ab} = P(Y^* = a | Y= b)$ denote the outcome misclassification parameters with $a$ and $b \in \{0, 1\}$. Thus, $p_{10} = P(Y^*=1 | Y=0)$ denotes one minus the specificity of the error prone outcome and  $p_{11} = P(Y^*=1 | Y=1)$ is its sensitivity. Using the validation sample, we can obtain estimates of the misclassification probabilities, denoted by $\hat{p}_{10}$ and $\hat{p}_{11}$. Shu and Yi \cite{shu2019causal} demonstrated that under the assumption of homogeneous misclassification probabilities (i.e., $P(Y^* = a | Y=b, X, T=t) = P(Y^* = a | Y=b)$) and assuming that ${p}_{11} \neq {p}_{10}$, a consistent estimator of $\tau$ can be expressed using $\hat{p}_{10}$ and $\hat{p}_{11}$. 

Ultimately, Shu and Yi \cite{shu2019causal} propose estimating the ATE with a weighted combination of the ATE estimate from the validation sample alone, $\hat{\tau}(Y_\mathcal{V}, \cdot)$, and the ATE estimate obtained from non-validation individuals, $\hat{\tau}(\cdot, Y^*_{\mathcal{V}^{C}}$). The forms of these estimators are reproduced below using our notation:

\begin{equation} \label{ShuYiEstimators}
\begin{split}
& \hat{\tau} (Y_\mathcal{V}, \cdot) = \frac{1}{n_{\mathcal{V}}}\sum_{i=1}^n V_i\frac{T_i Y_i}{\hat{e_i}} - \frac{1}{n_{\mathcal{V}}}\sum_{i=1}^n V_i \frac{(1-T_i) Y_i}{(1-\hat{e_i})}\\
&\hat{\tau}(\cdot, Y^*_{\mathcal{V}^{C}}) = \frac{1}{\hat{p}_{11} - \hat{p}_{10}} \left \{\frac{1}{n - n_{\mathcal{V}}}\sum_{i=1}^n (1-V_i) \frac{T_i Y^*_i}{\hat{e_i}} - \frac{1}{n-n_{\mathcal{V}}}\sum_{i=1}^n (1-V_i) \frac{(1-T_i) Y^*_i}{(1-\hat{e_i})} \right \} \\
&\hat{\tau} (Y_\mathcal{V}, Y^*_{\mathcal{V}^C}) = \frac{wn_{\mathcal{V}}}{wn_{\mathcal{V}} + (1-w)(n-n_{\mathcal{V}})}\hat{\tau} (Y_\mathcal{V}, \cdot) + \left\{ 1-\frac{wn_{\mathcal{V}}}{wn_{\mathcal{V}} + (1-w)(n-n_{\mathcal{V}})} \right\} \hat{\tau}(\cdot, Y^*_{\mathcal{V}^{C}})
\end{split}
\end{equation}
 
\noindent The weight $w$, $0 \leq w \leq 1$, is typically set as $w=0.5$, which weights the validation and non-validation samples proportional to their sample sizes. The weight $w$ can also be selected optimally to achieve the most efficient estimator amongst all estimators with the same form (details given in Shu and Yi \cite{shu2019causal}). 

Unlike the IPW estimators from Section 2.1, these estimators are appropriate in the presence of measurement error when the validation sample is a SRS. However, as highlighted in the introduction, the validation sample for the ACT study is a non-probability sample. Therefore, we must consider alternative estimators for the ATE when using the ACT data.
 
\subsection{Handling Outcome Misclassification in the EHR sample and Selection Bias in the Validation Sample}

We propose alternative estimators to account for  selection bias in the validation sample while simultaneously addressing misclassification in the outcomes in the non-validation data. One option is to revise Shu and Yi's estimator (i.e., $\hat{\tau} (Y_\mathcal{V}, Y^*_{\mathcal{V}^C})$) to incorporate validation sample selection propensities. We also assume that the mechanism for misclassification in the outcome is the same in the full EHR sample and the non-validation sample. 
We denote the probability of being selected into the validation sample by $\pi_{V}(T, X) = P(V=1 | T, X)$, where estimates of this quantity are denoted as $\hat{\pi}_V$, dropping the dependence on $T$ and $X$ to simplify notation. We make the following additional assumptions: (1) the model for $\pi_{V}$ is known and correctly specified; (2) positivity, i.e., $0 < \pi_V(T, X) < 1$; and (3) conditional ignorability of selection, i.e.,  $V \perp (Y_0, Y_1) | T, X$. Under standard $M$-estimator regularity conditions \cite{wooldridge2002econometric, wooldridge2002inverse}, the newly proposed estimators can be shown to be consistent estimators of the ATE by writing them in the form of an M-estimator which includes estimating equations for the parameters of both the treatment and selection propensity score models. Furthermore, we assume the misclassification probabilities $p_{11}$ and $p_{10}$ are homogeneous, i.e., independent of $X$ and $T$, as in Shu and Yi.\cite{shu2019causal} Define $\hat{\tau}^S (Y_{\mathcal{V}}, \cdot)$ and $\hat{\tau}^S (\cdot, Y^*_{\mathcal{V}^C})$ as follows:

\begin{equation} 
\openup 3\jot
\begin{split}
\label{PreShuYiSelect}
\hat{\tau}^S(Y_{\mathcal{V}}, \cdot) &= \frac{1}{n}\sum_{i=1}^n V_i\frac{T_i Y_i}{\hat{e_i}\hat{\pi}_{V,i}} - \frac{1}{n}\sum_{i=1}^n V_i\frac{(1-T_i) Y_i}{(1-\hat{e_i})\hat{\pi}_{V,i}} \\
\hat{\tau}^S(\cdot, Y^*_{\mathcal{V}^C}) &= \left (\sum_{i=1}^n \frac{(1 - V_i)T_i}{\hat{e_i}(1-\hat{\pi}_{V,i}) }\right)^{-1}\left(\sum_{i=1}^n \frac{(1 - V_i)T_i Y_i^*}{\hat{e_i}(1-\hat{\pi}_{V,i}) }\right) \\
&- \left(\sum_{i=1}^n \frac{(1 - V_i)(1-T_i)}{(1-\hat{e_i})(1-\hat{\pi}_{V,i}) }\right)^{-1} \left(\sum_{i=1}^n \frac{(1 - V_i)(1-T_i)Y_i^*}{(1-\hat{e_i})(1-\hat{\pi}_{V,i}) }\right). \\
\end{split}
\end{equation}

\noindent Note that we adopt the Hajek form \cite{hajek1971comment} of the IPW estimate of the ATE arising from the non-validation individuals in equation (\ref{PreShuYiSelect}) to improve efficiency. 

We first propose an estimator as a direct extension of Shu and Yi's denoted by $\hat{\tau}^S (Y_\mathcal{V}, Y^*_{\mathcal{V}^C})$ that combines the two estimators above, $\hat{\tau}^S (Y_\mathcal{V}, \cdot)$ and $\hat{\tau}^S (\cdot, Y^*_{\mathcal{V}^C})$. We use $w=0.5$ to weight the first term in equation (\ref{ShuYiSelect}) below by the size of the validation sample.\cite{shu2019causal} Our proposed estimator is defined as follows:
\begin{equation}
\openup 3\jot\label{ShuYiSelect}
\begin{split}
\hat{\tau}^S(Y_{\mathcal{V}}, Y^*_{\mathcal{V}^C}) &= \frac{n_{\mathcal{V}}}{n}\hat{\tau}^S(Y_{\mathcal{V}}, \cdot)
+ \frac{n - n_{\mathcal{V}}}{n}\left(\frac{1}{\hat{p}_{11} - \hat{p}_{10}}\right) \hat{\tau}^S(\cdot, Y^*_{\mathcal{V}^C}).
\end{split} 
\end{equation}

In addition to proposing $\hat{\tau}^S (Y_\mathcal{V}, Y^*_{\mathcal{V}^C})$ which incorporates sample selection propensities, we newly consider an alternative  estimator that incorporates sample selection propensities and leverages \textit{all} silver-standard outcomes, rather than only those coming from individuals in the non-validation sample. Let $\hat{\tau}(\cdot, Y^*) $ denote an estimator of the ATE that incorporates misclassification probabilities from the validation data and all silver-standard outcomes,
\begin{equation} \label{allSilver}
\hat{\tau}(\cdot, Y^*) =  \Bigl(\frac{1}{\hat{p}_{11} - \hat{p}_{10}}\Bigr) \Bigl\{ \Bigl( \sum_{i=1}^n \frac{T_i}{\hat{e_i} } \Bigr)^{-1} \Bigl(\sum_{i=1}^n \frac{T_i Y_i^*}{\hat{e_i} }\Bigr)  - \Bigl(\sum_{i=1}^n \frac{(1-T_i) }{(1-\hat{e_i}) }\Bigr)^{-1}\Bigl(\sum_{i=1}^n \frac{(1-T_i) Y_i^*}{(1-\hat{e_i}) }\Bigr) \Bigr\}.\\
\end{equation}

\noindent As an alternative to $\hat{\tau}^S (Y_{\mathcal{V}}, Y^*_{\mathcal{V}^C})$, we propose the following estimator, which is a weighted combination of  $\hat{\tau}^S (Y_{\mathcal{V}}, \cdot)$ and $\hat{\tau}(\cdot, Y^*)$:  
\begin{equation} 
\hat{\tau}^S(Y_{\mathcal{V}}, Y^*) = b\hat{\tau}^S(Y_{\mathcal{V}}, \cdot) + (1-b)\hat{\tau}(\cdot, Y^*)
\end{equation}

\noindent An optimal choice of $b$ can be derived to minimize the variance of an estimator of the form of $\hat{\tau}^S(Y_{\mathcal{V}}, Y^*)$. In addition to requiring $0 \leq b \leq 1$, we enforce the following constraint: 
\begin{equation*}
Var(\hat{\tau}^S(Y_{\mathcal{V}}, \cdot)) + Var(\hat{\tau}(\cdot, Y^*)) - 2Cov(\hat{\tau}^S(Y_{\mathcal{V}}, \cdot), \hat{\tau}(\cdot, Y^*)) \geq 0.
\end{equation*}

\noindent Then following the same logic as Shu and Yi \cite{shu2019causal}, it can be shown that the weight that minimizes $Var(\hat{\tau}^S(Y_{\mathcal{V}}, Y^*))$ is 
\begin{equation} 
\label{wopt}
b_{opt} = \frac{Var(\hat{\tau}(\cdot, Y^*)) - Cov(\hat{\tau}^S(Y_{\mathcal{V}}, \cdot),\hat{\tau}(\cdot, Y^*))}{Var(\hat{\tau}^S(Y_{\mathcal{V}}, \cdot)) + Var(\hat{\tau}(\cdot, Y^*)) - 2Cov(\hat{\tau}^S(Y_{\mathcal{V}}, \cdot), \hat{\tau}(\cdot, Y^*))}.
\end{equation}

\noindent Let the estimator that incorporates $b_{opt}$ be denoted by $\hat{\tau}^S_{opt}(Y_{\mathcal{V}}, Y^*)$. Thus,$\hat{\tau}^S_{opt}(Y_{\mathcal{V}}, Y^*) = b_{opt}\hat{\tau}^S (Y_{\mathcal{V}}, \cdot) + (1-b_{opt})\hat{\tau}(\cdot, Y^*)$. 

To summarize, $\hat{\tau} (Y_{\mathcal{V}}, \cdot)$ and $\hat{\tau}^S (Y_{\mathcal{V}}, \cdot)$ use only information from the validation data without or with adjustment for selection bias, respectively. $\hat{\tau}(\cdot, Y^*)$ uses all silver-standard outcomes  but only  validation data to estimate the misclassification parameters, $p_{10}$ and $p_{11}$. The estimators that integrate information from the silver-standard and validation data are $\hat{\tau} (Y_{\mathcal{V}}, Y^*_{\mathcal{V}^C})$, $\hat{\tau}^S (Y_{\mathcal{V}}, Y^*_{\mathcal{V}^C})$, $\hat{\tau}^S(Y_{\mathcal{V}}, Y^*)$, and $\hat{\tau}^S_{opt}(Y_{\mathcal{V}}, Y^*)$. The estimators that account for selection bias by incorporating the propensity of being selected into the validation sample are $\hat{\tau}^S (Y_{\mathcal{V}}, \cdot)$, $\hat{\tau}^S (Y_{\mathcal{V}}, Y^*_{\mathcal{V}^C})$, $\hat{\tau}^S(Y_{\mathcal{V}}, Y^*)$, and $\hat{\tau}^S_{opt}(Y_{\mathcal{V}}, Y^*)$. 


\subsection{Inference}
Standard errors (SE) were estimated via a stacked estimating equation framework. We defined and stacked unbiased estimating equations for parameters of the treatment assignment model, parameters of the misclassification models, and a given estimator for the ATE. If relevant to the estimator, we also included an unbiased estimating equation for parameters of the validation sample selection model. We then solved these estimating functions and estimated the covariance matrix with an empirical sandwich estimator. Subsequent variance estimates were used to construct 95\% confidence intervals and estimate coverage. Additional details are contained in the Supplement. An R package, \texttt{validatEHR} for implementing the proposed methods is available via GitHub (https://github.com/jshen650/validatEHR).

\section{Simulation Studies}
\label{s:simulations}

We conducted simulation studies to compare the performance of existing estimators and our newly proposed estimators for the ATE with binary outcomes subject to misclassification, assuming validation data are available. We investigated performance when the validation sample was a simple random sample (SRS) as well as when the validation sample was a non-probability sample. Although not shown here, we also studied these estimators under a range of alternative conditions, such as varying outcome misclassification rates and sizes of the validation sample; varying the magnitude of validation sample selection bias; assuming heterogeneous misclassification probabilities; and misspecifying the validation sample selection model. Full details for these additional scenarios are provided in the Supplement.

\subsection{Data Generation}
Let $\mathbf{X} = (X_1, X_2, X_3, X_4, X_5)^{T}$ denote a $5 \times 1$ vector of baseline variables, and let $expit(u)$ denote the inverse of the logit function. Let $1_{d}$ denote a $d\times 1$ vector of 1s. We generated the complete data, $\{(\mathbf{X}_{i}, T_{i}, Y_{i}, Y^{*}_{i})\}_{i=1}^{n}$, independently for each individual as follows:
\begin{equation} \label{genDat}
\begin{split}
&\mathbf{X}_{i} \sim MVN(0_{5 \times 1}, I_{5\times 5}), \hspace{5mm} \pi_{T}(\mathbf{X}_{i}) = expit(0.8 + 0.3(1_{5}^T\mathbf{X}_{i})),\\
&T_{i} \sim Ber(\pi_{T}(\mathbf{X}_{i})), \hspace{5mm} \pi_{Y}(T_{i},\mathbf{X}_{i}) = expit(-3.9 + T_{i} + 1_{5}^T\mathbf{X}_{i}),\\ 
&Y_{i} \sim Ber(\pi_{Y}(T_{i},\mathbf{X}_{i})) 
\end{split}
\end{equation}
Values of $\bm{Y}^*$ were simulated based on pre-specified values of $p_{11}$ and $p_{10}$. We assume that all $n$ individuals have the observed information $(\mathbf{X}, T, Y^*)$. A subset of these $n$ individuals comprise the validation sample $\mathcal{V}$ of size $n_{\mathcal{V}}$ and collectively contribute information $\{(\mathbf{X}_{i}, T_{i}, Y_{i}, Y^{*}_{i})\}_{i \in \mathcal{V}}$. The prevalences of $Y^*, Y, \text{and } T$ in our simulated data sets were targeted to reflect their prevalences in the ACT study. For the SRS validation samples, $P(Y^*=1) = 0.3, P(Y=1) = 0.14$, and $P(T=1)=0.67$. For the non-probability validation samples, $P(Y^*=1) = 0.4, P(Y=1) = 0.37$, and $P(T=1)=0.84$.

Samples of size $n=5000$ were simulated following the methods described previously. Simulated validation samples were nested within the full sample. We simulated a random variable $V_{i} \sim Ber(\pi_V(\mathbf{X}_i, T_{i}))$ to determine selection of individual $i$ into $\mathcal{V}$. We consider two types of validation samples: simple random samples (SRS) and non-probability samples. For validation samples that were SRS, we define $\pi_V = n_{\mathcal{V}}/n$. For validation samples that were non-probability samples, we defined $\widetilde{X}_{i} = (1, T_{i}, \mathbf{X}_{i}^{T})^T$ and $\alpha_{0} = (\alpha_{intercept}, 0.5, 1, 1, 1, 1, 0)$. Then $\pi_V(\mathbf{X}_{i}) = expit(\alpha_{0}^T\widetilde{X}_{i})$, where the choice of $\alpha_{intercept}$ varies to achieve targeted values for $n_{\mathcal{V}}$. To simulate validation samples of similar size to that of the ACT study, we targeted $n_{\mathcal{V}} = 850$ which is 17\% of the full sample, $n$. Based on characteristics of the motivating real data example, we used misclassification probabilities of $p_{11} = 0.67$ and $p_{10} = 0.24$.

An estimate of the true ATE was obtained by generating large data sets ($n=50,000$) from the true model, calculating the IPW estimate of the ATE, and taking the average of this process across 5,000 iterations. Using each of the estimators described in Section \ref{s:methods}, we estimated the ATE and its sandwich standard error (SE). We then constructed approximate 95\% confidence intervals (CI) and assessed coverage for the estimate of the true ATE. We ran 5,000 simulation iterations for each simulation scenario. Preliminary simulations indicated that using the value for $w$ that minimizes the variance of $\hat{\tau} (Y_\mathcal{V}, Y^*_{\mathcal{V}^C})$ led to biased estimates in order to achieve optimal efficiency. Thus, we proceeded with using $w=0.5$ for $\hat{\tau} (Y_\mathcal{V}, Y^*_{\mathcal{V}^C})$.


\subsection{Results}
\begin{figure}[!htb]
  \centering
  \includegraphics[width=1\linewidth,scale = 1]{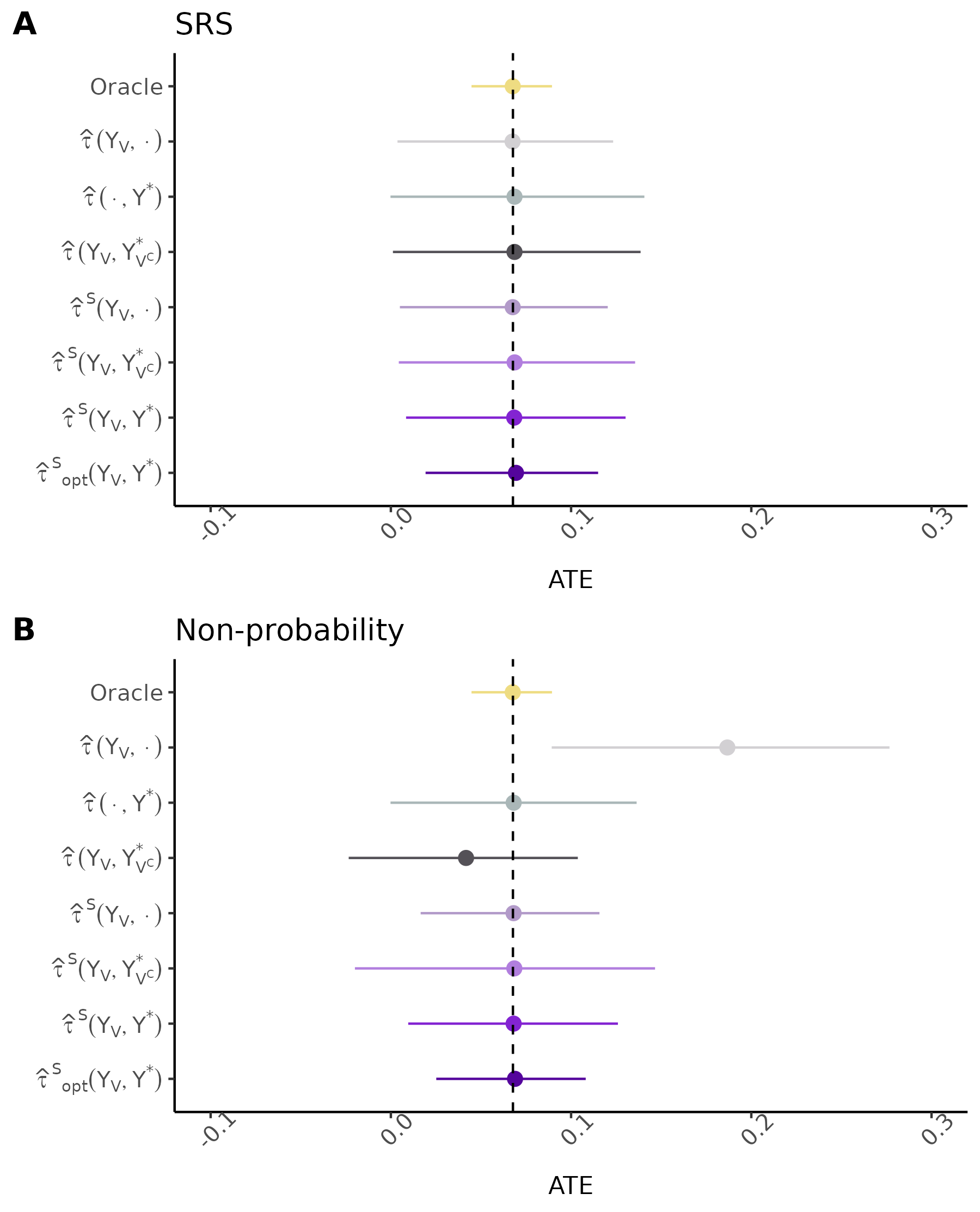}
  \centering
  \caption{Empirical 95\% CIs for all estimators when validation samples are A) SRS or B) non-probability samples. Results are based on 5,000 simulation iterations. The total sample size was $n=5000$, and the validation sample size was approximately $n_{\mathcal{V}} \approx 850$. The dashed line is the true ATE. Estimators that account for sample selection propensities for estimating the ATE are indicated by the form $\hat{\tau}^S$. Otherwise, $\hat{\tau}$ denotes estimators that do not account for validation sample selection propensities.}
  \label{fig:simres_both}
\end{figure}

Point estimates and 95\% empirical confidence intervals for all estimators for both validation sample types (e.g. SRS and non-probability) are shown in Figure \ref{fig:simres_both} for validation samples of size $n_{\mathcal{V}} \approx 850$. The black line represents our estimate of the true ATE, which was approximately 0.07. Corresponding estimates of the bias, average empirical standard error, average sandwich standard error, and confidence interval coverage probabilities are provided in Table \ref{tab:simres_both}. Obtaining results from 5000 iterations for our proposed estimators took less than 45 seconds for a given estimator when parallelizing over 100 cores on the Penn Medicine Academic Computing Services High Performance Cluster.

As shown in the left panel of Figure \ref{fig:simres_both}, when the validation samples were SRS, all estimators were unbiased with similar standard errors. Although $\hat{\tau}(Y_{\mathcal{V}},\cdot)$ had the largest standard errors, incorporating additional information from $\bm{Y}^*$ produced relatively modest efficiency gains due to the substantial mislassification in $\bm{Y}^*$. Coverage of 95\% confidence intervals was approximately nominal across all estimators (Table \ref{tab:simres_both}). Differences between average empirical standard error estimates and average sandwich standard error estimates were quite small, indicating that the sandwich standard error estimates were close to the true standard errors. Our newly proposed estimators $\hat{\tau}^S(Y_\mathcal{V}, Y^*)$ and $\hat{\tau}^S_{opt}(Y_\mathcal{V}, Y^*)$ improved efficiency relative to $\hat{\tau} (Y_\mathcal{V}, Y^*_{\mathcal{V}^C})$. The differences in the average sandwich SEs of $\hat{\tau}^S (Y_\mathcal{V}, Y^*)$ and $\hat{\tau}^S_{opt}(Y_\mathcal{V}, Y^*)$ compared to that of $\hat{\tau} (Y_\mathcal{V}, Y^*_{\mathcal{V}^C})$ were 0.002 and 0.012, respectively. The most efficient estimator was $\hat{\tau}^S_{opt}(Y_\mathcal{V}, Y^*)$, which was expected since the choice of weight $w$ prioritizes lower variance. 

When validation samples were non-probability samples, estimators that failed to account for selection into the validation sample had high bias. As seen in the right panel of Figure \ref{fig:simres_both}, $\hat{\tau} (Y_\mathcal{V}, \cdot)$ and $\hat{\tau}^S (Y_\mathcal{V}, Y^*_{\mathcal{V}^C})$ were centered away from the true ATE. In Table \ref{tab:simres_both}, the magnitude of the bias for these estimators exceeded 0.02 or about 30\% of the true ATE. While $\hat{\tau}^S (Y_\mathcal{V}, Y^*_{\mathcal{V}^C})$ adjusts for selection into the validation sample, this estimator was not as efficient as estimators that incorporated data on $\bm{Y}^*$ from all subjects rather than a subset (i.e., $\hat{\tau}^S(Y_\mathcal{V}, Y^*)$ and $\hat{\tau}^S_{opt}(Y_\mathcal{V}, Y^*)$). Our newly proposed estimators $\hat{\tau}^S(Y_\mathcal{V}, Y^*)$ and $\hat{\tau}^S_{opt}(Y_\mathcal{V}, Y^*)$ were both unbiased and among the most efficient in this scenario. Average empirical standard error estimates for all estimators were similar to corresponding average sandwich standard error estimates, once again indicating accurate estimation of standard errors. Results were similar across simulations that varied the misclassification rates and validation sample sizes (see Supplement).

\section{Real Data Analysis}
\label{s:realdata}

We used data from the Adult Changes in Thought (ACT) study to estimate the ATE of hypertension on development of Alzheimer's disease (AD) neuropathology using existing and our newly proposed estimators. Since 1994, the Adult Changes in Thought (ACT) Study has recruited participants from random samples of Kaiser Permanente Washington health plan members who are at least 65 years of age, dementia-free, do not reside in a nursing home, and have been enrolled in the health plan for at least 2 years. Information pertaining to demographic characteristics, medical history, and functional status was assessed at baseline and follow-up interviews occurring every 2 years.\cite{kukull2002dementia} Hypertension has previously been observed to increase the risk of dementia and AD.\cite{sierra2020hypertension, lennon2019midlife, decarli2021link} Previous studies with data from ACT have observed associations between hypertension and clinical dementia \cite{li2007age} and between systolic blood pressure and certain neuropathologic measures of AD.\cite{wang2009blood} In the full ACT cohort, Li et al.\cite{li2007age} found that higher systolic blood pressure was associated with greater dementia risk in participants aged 65-74 years old. And in participants who were 65-80 years old included in the autopsy sample, Wang et al.\cite{wang2009blood} found that systolic blood pressure was associated with occurrence of cerebral microinfarcts, a neuropathologic AD measure. Both the ACT cohort and autopsy sample have notably increased in size since these earlier studies, however. Presently, the ACT cohort is more than twice the size of the cohort studied in Li et al.\cite{li2007age}, and the autopsy sample is more than three times the size of the autopsy sample studied in Wang et al.\cite{wang2009blood} We therefore investigated the relationship between hypertension and AD neuropathology using updated data from the ACT study. 

We assumed that individuals in the ACT cohort represent a simple random sample of members of the Kaiser Permanente Washington health system. Silver-standard outcome data from clinical evaluation for AD is available for all members of this cohort. Gold-standard outcome data in the form of AD neuropathology ascertained from autopsy is available for a subset of participants who consented. Following the approach of Sonnen et al.\cite{sonnen2009neuropathology}, presence of AD neuropathology was classified based on Braak stage and Consortium to Establish a Registry for Alzheimer's Disease (CERAD) ratings. The CERAD rating is a measure of neuritic plaques, ranging from absent (0) to frequent (3), where greater frequency of neuritic plaques indicates AD.\cite{mirra1991consortium} Braak stage ranges from Stage 0 to Stage VI based on severity of neurofibrillary tangles, where Stage VI is the most severe.\cite{braak1991neuropathological} Individuals with Braak stage $\geq V$ and CERAD rating $\geq$ 2 were classified as having AD neuropathology (i.e., $Y=1$).  

 An individual was classified as hypertensive if their maximum observed systolic blood pressure value across all longitudinal clinical visits was $\geq 140$ mmHg. The propensity model for hypertension included a binary variable assessing usage of hypertension medication (e.g., ever vs never); body mass index (BMI) at baseline; age at last study visit (quintiles); race (White or non-White); and gender (female vs male). These variables were included given their known associations with hypertension and AD in order to support the assumption of conditional exchangeability of exposure for estimating the ATE. BMI, age, race, and gender, for instance, have all been observed to be risk factors for hypertension \cite{slama2002prevention} and AD \cite{moody2021body, armstrong2019risk, lennon2022black, podcasy2016considering}, and associations between hypertension medication usage and AD incidence have also been noted in previous studies.\cite{yasar2013antihypertensive, barthold2018association} 

For the autopsy sample selection model, we included variables previously identified by the ACT study as being associated with inclusion in the autopsy cohort.\cite{haneuse2009adjustment} These included ACT study cohort (3 levels: original, expansion, and recruitment), clinical dementia status at final study visit, age at last study visit (quintiles), race, gender, and hypertension. An important assumption for adjusting for autopsy sample selection bias here is that all components of the exposure and other covariates are observable on all individuals, including those not selected (i.e., missing-at-random, or MAR, assumption). This assumption cannot be verified from the data alone and follows instead from scientific reasoning and precedents set by other analyses; subsequently we included variables relevant for autopsy sample selection that were identified in Haneuse et al.\cite{haneuse2009adjustment} Some missingness existed in the data, and we excluded 94 individuals with missing covariate data. Thus, we included $n=5669$ individuals from the full ACT cohort, where $n_{\mathcal{V}}=837$ of these individuals were part of the autopsy subsample.

The prevalence of the clinical diagnoses of AD ($\bm{Y}^*$) and hypertension ($T$) were greater in the autopsy cohort compared to the full ACT sample (Table \ref{tab:ACTSummary}). \noindent From the full ACT cohort, $P(Y^* = 1) = 0.19$, while in the autopsied cohort,  $P(Y^* = 1) = 0.38$. For hypertension, $P(T = 1) = 0.69$ in the full ACT sample and $P(T = 1) = 0.76$ in the autopsied subsample. From the autopsied cohort, $P(Y=1) = 0.32$. Compared to the full cohort, the autopsy cohort was generally older and sicker. This difference between the autopsy cohort and the full cohort may be expected, given that inclusion into the autopsy cohort requires death. As mentioned earlier, inclusion into the autopsy cohort required consent of participants. These consent rates have been noted to differ significantly between white and non-white individuals,\cite{haneuse2009adjustment} where a greater rate of consent is reflected in the higher proportion of white individuals represented in the autopsy cohort. The estimated sensitivity ($p_{11}$) and specificity ($p_{00}$) for $\bm{Y}^*$ were $\hat{p}_{11}=0.67$ and $\hat{p}_{00}=0.76$, respectively. A summary of all relevant variables for our analysis can be found in Table \ref{tab:ACTSummary}. Coefficient estimates of the treatment propensity and autopsy sample selection models can be viewed in the Supplement. 

\begin{figure}[!htb]
  \centering
  \includegraphics[width=1\linewidth,scale = 1]{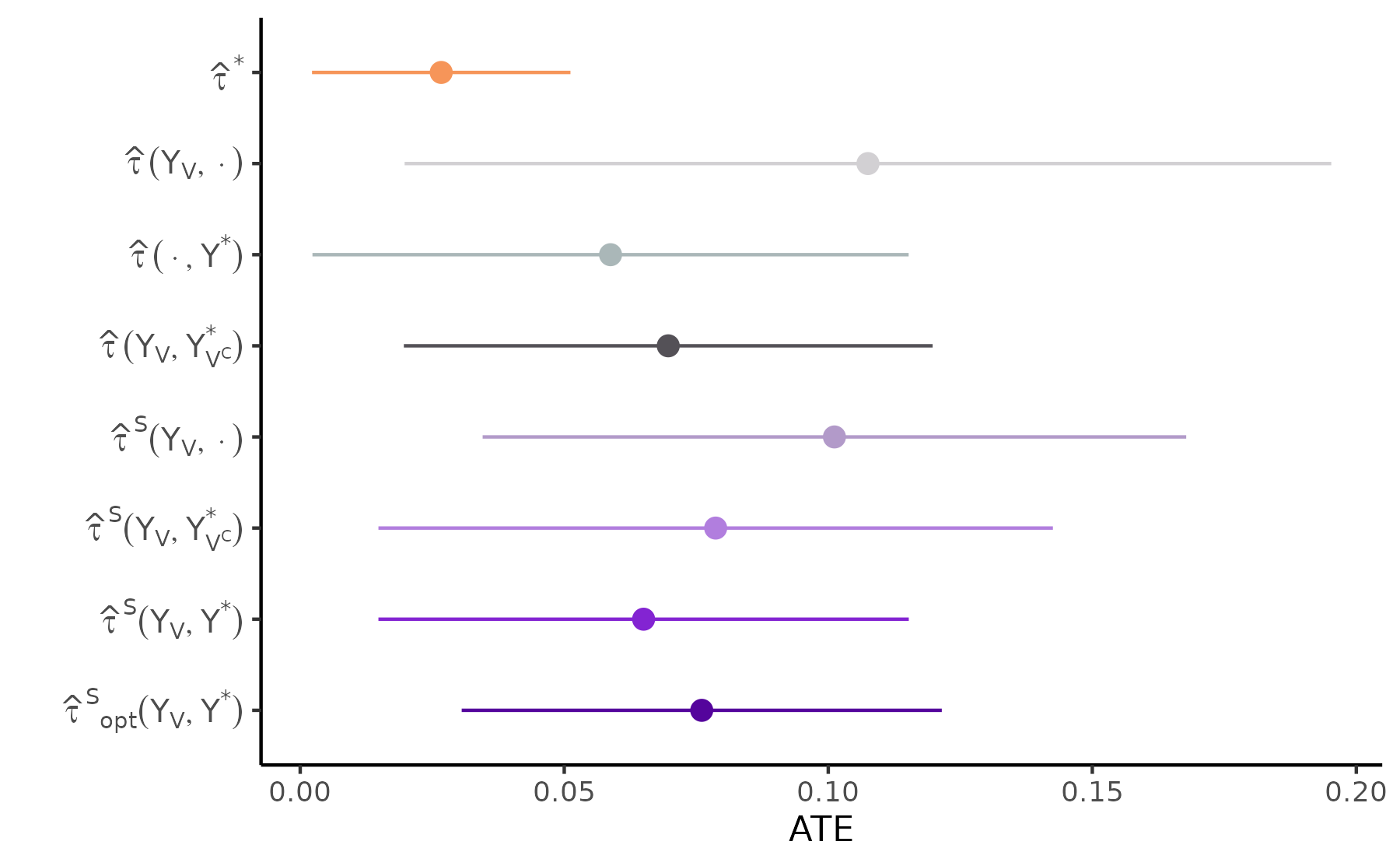}
  \centering
  \caption{Comparison of 95\% CIs based on all estimators using the ACT data. Estimators that account for sample selection propensities for estimating the ATE are indicated by the form $\hat{\tau}^S$. Otherwise, $\hat{\tau}$ denotes estimators that do not account for validation sample selection propensities.The naive estimate of the ATE, $\hat{\tau}^*$, uses only the silver-standard AD diagnoses as the outcomes. }
  \label{fig:act_est}
\end{figure}

Point estimates and 95\% confidence intervals of all estimators are shown in Figure \ref{fig:act_est}. \noindent In addition to the estimators described previously, we included a naïve estimate of the ATE based on using all silver-standard outcomes without adjustment for misclassification. Relative to other approaches, the naive estimator was substantially attenuated towards the null. All other estimators exhibited a positive estimate of the ATE, ranging from a 3\% to 8\% increase in the risk of AD neuropathology for individuals with hypertension. The point estimates are all slightly different, but these differences could plausibly be attributed to random variability. In this data set, accounting for sample selection appears to have little impact on the results. Both $\hat{\tau} (Y_\mathcal{V}, \cdot)$ and $\hat{\tau}^S (Y_\mathcal{V}, \cdot)$, which use only validation data, had the highest variance. Other methods that use silver-standard outcomes appeared to be more efficient (see results for: $\hat{\tau} (Y_\mathcal{V}, Y^*_{\mathcal{V}^C})$, $\hat{\tau}^S (Y_\mathcal{V}, Y^*_{\mathcal{V}^C})$, $\hat{\tau}^S(Y_\mathcal{V}, Y^*)$, and $\hat{\tau}^S_{opt}(Y_\mathcal{V}, Y^*)$). Furthermore, methods that use $(Y_\mathcal{V}, Y^*)$ were more efficient than those that used only $(Y_\mathcal{V}, Y^*_{\mathcal{V}^C})$.

\section{Discussion}
\label{s:discussion}

In this work, we considered the case of a large EHR-derived cohort augmented with gold-standard data from a non-probability validation sub-sample. Accurate estimation of the ATE relies on addressing potential outcome misclassification in the EHR-derived sample and potential selection bias in the validation sample. Here, we presented estimators for the ATE that correct for misclassification in the outcome and selection bias of the validation sample simultaneously. In simulations, we found that existing estimators and our proposed estimators were all unbiased for the true ATE when the validation sample was an SRS. However, when the validation sample was a non-probability sample, estimators that failed to account for selection were biased. In this setting, our proposed estimators – specifically, $\hat{\tau}^S(Y_\mathcal{V}, Y^*)$ and $\hat{\tau}^S_{opt}(Y_\mathcal{V}, Y^*)$ -- reduced bias and increased efficiency, improving inference for the ATE. In the ACT data analysis, we estimated the ATE using all estimators and found that failing to correct for misclassification in the outcome led to substantial attenuation to the null. Broadly, the relative magnitude of confounding, misclassification, and selection propensity will influence bias. In our context, estimates accounting for selection bias were relatively similar to those that did not incorporate selection propensities. When comparing Row 1 (orange) to Row 3 (medium silver) in Figure \ref{fig:act_est}, we see that attenuation to the null is being driven by the presence of outcome misclassification in the ACT data. 
  Augmenting EHR with thoughtful prospective data collection can overcome limitations inherent in EHR or other secondary data sources that are not collected for research purposes. Leveraging validation samples may lead to improved estimation and inference in downstream models, benefiting clinical knowledge and outcomes.\cite{zhou2007efficient, amorim2021two} While validation samples can be carefully collected such that every individual has known probability of being selected (i.e., a probability sample), not all scenarios allow for this possibility. An advantage of our methods is that they allow for use of a non-probability validation sample which may represent a convenience sample or population sub-sample wiling to consent to research participation. Importantly, our methods require a correctly specified selection model, i.e., that all factors that explain inclusion in the non-probability validation sample are measured and included the selection model.


Choice of the treatment propensity model and validation sample selection models can also impact the effectiveness of estimators that address both misclassification in the outcome and sample selection bias. While misspecification of the treatment propensity and autopsy sample selection models could impact the accuracy of ATE estimation, we did not pursue studies of potential misspecification of these models (e.g., developing accompanying sensitivity analyses) since our focus for the ACT data analysis was on comparing the estimators. In supplemental simulations, our estimators demonstrated moderate robustness to selection model misspecification (Supplementary Table 7). We observed in the ACT data analysis that estimators that corrected for both misclassification in the outcome and selection bias in the validation sample performed comparably to estimators that only corrected for misclassification in the outcome. In this example, accounting for sample selection bias may play a relatively small role relative to the magnitude of bias due to misclassification. The effect of hypertension on the probability of inclusion in the autopsy sample, either through increased probability of death or willingness to consent to autopsy, may be fully mediated by other variables included in the treatment propensity model such as age and treatment use of anti-hypertensive medications, rendering approaches to address selection bias unnecessary. 

The generalizability of results depends on how the target population is defined with respect to the EHR. Patients in a given regional healthcare system may differ notably from the general population in demographics such as age, sex, or race as well as other variables related to health and social determinants of health.\cite{casey2016using} Particularly, if the target population is the total population in a given region as opposed to those served by a specific healthcare system, certain subpopulations may be over- or under-represented relative to the regional population. Augmenting EHR with higher quality data from other sources, such as patient registries, can potentially improve generalizability while also mitigating EHR data quality concerns such as missingness.\cite{ehrenstein2019obtaining} Relative to relying on one collection of EHR data, bringing together multiple data sources can provide a more comprehensive, accurate, and timely presentation of patient statuses and medical histories thereby offering greater insight into characteristics of the target population.\cite{weiskopf2019towards, ehrenstein2019obtaining} 


While we focused exclusively on IPW estimation of the ATE with binary outcomes, future work could consider how to adapt our estimators to a range of additional estimation approaches and settings. While we addressed correcting for misclassification in the outcome, accounting for measurement error in predictors could also be of interest in the ACT study and other contexts. Potential areas for additional methods development include formulating multiply robust versions of our estimators, modifying our method for estimands such as the conditional average treatment effect (CATE), or considering continuous outcomes instead of binary outcomes. Developing and comparing multiply robust estimators of the ATE to adjust for misclassification in the outcome while modeling selection into the validation sample would be an exciting extension which merits significant adaptation of our estimators and additional evaluations of comparative performance. A supplemental exploration showed biased performance of estimators described in this paper under a misspecified validation sample selection, highlighting the need for further development in this area. Such extensions would require multiply robust forms of our estimators that would be robust to misspecification of (one of) the outcome, treatment propensity, and validation sample selection propensity models. Another consideration in our analysis of the ACT data was how to address individuals with missing covariate data. Following previous analyses of the ACT data,\cite{haneuse2009adjustment} we chose to exclude individuals with missing covariate information. However, concerns over potential bias arising from missingness that is not completely random are warranted. Although outside the scope of this work, exploring the impacts of different missingness mechanisms in covariates and incorporating strategies to mitigate the impacts (e.g., multiple imputation, inverse probability weighting) would constitute another important extension of our work. In conclusion, when gold-standard measures are not available for large EHR-derived samples but resources are available to obtain gold-standard measures from a validation sample, estimators that integrate both sets of outcomes improve inference for the ATE. Particularly when the validation sample is a non-probability sample, our proposed estimators reduce bias and increase efficiency, supporting valid and efficient inference using EHR data. 

\bibliographystyle{unsrtnat}
\bibliography{WileyNJD-AMA.bib}

\clearpage

\section*{Tables}

\begin{table}[!htb]
    \caption{Simulation results for bias, average empirical SE, average sandwich SE, and 95\% confidence interval coverage probabilities. For results aside from oracle ($\hat \tau$), the lowest estimates of bias, empirical SE, and sandwich SE are bolded. Estimates of coverage that are closest to nominal also are bolded. Results are reported for scenarios in which the validation sample is an SRS or a non-probability sample. Our proposed estimators take the form $\hat{\tau}^S (Y_{\mathcal{V}}, \cdot)$.}
    \label{tab:simres_both}
    \centering
    \begin{tabular}{llcccc} 
    \hline
    \textbf{Validation} & \textbf{Estimator} & \textbf{Bias} & \textbf{Average} & \textbf{Average} & \textbf{Coverage} \\
    \textbf{Sample Type} &  & \ & \textbf{Empirical SE} & \textbf{Sandwich SE} & \\
    \hline
    SRS & $\hat{\tau}$ & 0.000 & 0.012 & 0.011 & 0.948 \\
    & $\hat{\tau} (Y_\mathcal{V}, \cdot)$ & \textbf{0.000} & 0.031 & 0.030 & 0.934 \\
    & $\hat{\tau} (Y_\mathcal{V}, Y^*_{\mathcal{V}^C})$ & 0.001 & 0.035 & 0.035 & 0.957 \\
    & $\hat{\tau}^S (Y_\mathcal{V}, Y^*_{\mathcal{V}^C})$ & 0.001 & 0.034 & 0.033 & \textbf{0.947} \\
    & $\hat{\tau}^S (Y_\mathcal{V}, \cdot)$ & \textbf{0.000} & 0.030 & 0.030 & 0.941 \\
    & $\hat{\tau}(\cdot, Y^*)$ & 0.001 & 0.036 & 0.035 & \textbf{0.949} \\
    & $\hat{\tau}^S(Y_\mathcal{V}, Y^*)$ & 0.001 & 0.031 & 0.030 & \textbf{0.947} \\
    & $\hat{\tau}^S_{opt}(Y_\mathcal{V}, Y^*)$ & 0.002 & \textbf{0.025} & \textbf{0.024} & 0.934 \\
    \hline
    Non-probability & $\hat{\tau}$ & 0.000 & 0.012 & 0.011 & 0.948 \\
    & $\hat{\tau} (Y_\mathcal{V}, \cdot)$ & 0.119 & 0.049 & 0.048 & 0.318 \\
    & $\hat{\tau} (Y_\mathcal{V}, Y^*_{\mathcal{V}^C})$ & -0.026 & 0.033 & 0.032 & 0.868 \\
    & $\hat{\tau}^S (Y_\mathcal{V}, Y^*_{\mathcal{V}^C})$ & 0.001 & 0.044 & 0.041 & \textbf{0.947} \\
    & $\hat{\tau}^S (Y_\mathcal{V}, \cdot)$ & \textbf{0.000} & 0.026 & 0.024 & 0.937 \\
    & $\hat{\tau}(\cdot, Y^*)$ & \textbf{0.000} & 0.036 & 0.035 & \textbf{0.951} \\
    & $\hat{\tau}^S(Y_\mathcal{V}, Y^*)$ & \textbf{0.000} & 0.030 & 0.030 & \textbf{0.949} \\
    & $\hat{\tau}^S_{opt}(Y_\mathcal{V}, Y^*)$ & 0.001 & \textbf{0.021} & \textbf{0.020} & 0.937 \\
    \hline
    \end{tabular}
\end{table}

\begin{table}[htb]
\caption{Summary of select characteristics for the full ACT cohort and the autopsy sub-sample which contains gold-standard AD diagnoses from individuals who consented to post-mortem brain autopsy.}
\label{tab:ACTSummary}
    \centering
    \begin{tabularx}{\textwidth}{X*{3}{>{\centering\arraybackslash}X}} 
    \hline
    Characteristic & Full ACT Cohort & Autopsy Sub-sample\\
    \hline
    n & 5669 & 837\\
    \hline
    Possible/Probable clinical AD diagnosis (\%) & 1102 (19.4) &  317 (37.9)\\
    \hline
    Previous or Current Hypertension Medication Use (\%) & 3685 (65.0) & 585 (69.9)\\
    \hline
    Hypertension (\%) & 3913 (69.0) & 632 (75.5)\\
    \hline
    Age Group at Last Visit (years) (\%) &  & \\
    
    [65,74] & 1131 (20.0) & 45 ( 5.4)\\
    
    (74,80] & 1332 (23.5) & 120 (14.3)\\
    
    (80,84] & 951 (16.8) & 153 (18.3)\\
    
    (84,89] & 1235 (21.8) & 244 (29.2)\\
    
    $>$89 & 1020 (18.0) & 275 (32.9)\\
    \hline
    White (\%) & 5078 (89.6) & 788 (94.1)\\
    \hline
    Female (\%) & 3276 (57.8) & 483 (57.7)\\
    \hline
    Any Dementia Diagnosis (\%) & 1333 (23.5) & 380 (45.4)\\
    \hline
    ACT Cohort (\%) &  & \\
    
    Original & 2567 (45.3) & 518 (61.9)\\
    
    Expansion & 785 (13.8) & 186 (22.2)\\
    
    Replacement & 2317 (40.9) & 133 (15.9)\\
    \hline
    BMI (mean (SD)) & 28.20 (5.25) & 27.78 (4.98)\\
    \hline
\end{tabularx}
\end{table}

\clearpage

\section*{Figure Captions} 
  
In addition to being included with individual figure files, figure captions also are reproduced here.

\clearpage

\section*{Acknowledgments}

This research was funded by the National Institute on Aging (U19AG066567, R21AG075574). Data collection for this work was additionally supported, in part, by prior funding from the National Institute on Aging (U01AG006781). All statements in this report, including its findings and conclusions, are solely those of the authors and do not necessarily represent the views of the National Institute on Aging or the National Institutes of Health. We thank the participants of the Adult Changes in Thought (ACT) study for the data they have provided and the many ACT investigators and staff who steward that data. You can learn more about ACT at: https://actagingstudy.org/

\subsection*{Data Availability Statement}

The data that support the findings of this study were provided with permission from ACT.

\subsection*{Conflict of interest}

The authors declare no potential conflict of interests.

\end{document}


\section*{Supplement for ``Integrating Misclassified EHR Outcomes with Validated Outcomes from a Non-probability Sample''}

\section{Estimation of Standard Errors for Proposed Estimators}

Similar to the approach proposed by Shu and Yi \cite{shu2019causal}, estimation of standard errors for the inverse probability-weighted (IPW) estimators of the average treatment effect (ATE) followed from defining and stacking unbiased estimating equations. These estimating equations pertained to parameters of the treatment model, parameters of the misclassification models, a given estimator for the ATE, and parameters of the validation sample selection model. Subsequent estimates of standard errors followed from solving estimating functions and estimating the covariance matrix with an empirical sandwich estimator. 

For $\hat{\tau}^S (Y_{\mathcal{V}}, Y^*_{\mathcal{V}^C})$, we construct estimating equations using the fact that the Hajek form of the IPW estimator of the ATE can be computed via weighted least squares \cite{aronow2017estimating}. Define:
\begin{equation}
\label{weightR}
R_i = \frac{(1-V_i)T_i}{e_i (1-\pi_{V,i})} + \frac{(1-V_i)(1-T_i)}{(1-e_i)(1-\pi_{V,i}))}
\end{equation}.

Let $\gamma$ represent parameters of the treatment model; $\eta$ represent parameters of the validation sample selection model; $\alpha$ and $\beta$ represent parameters of the weighted least squares regression model for $\bm{Y}^*$; $p_{11} = P(Y^* = 1 | Y=1)$; and $p_{10} = P(Y^*=1 | Y=0)$. Let $\theta$ include all model parameters associated with the estimator $\hat{\tau}^S (Y_{\mathcal{V}}, Y^*_{\mathcal{V}^C})$, we have that $\theta = (\tau^S (Y_{\mathcal{V}}, \cdot),\gamma^T, \eta^T ,\alpha, \beta, p_{11}, p_{10})^T$. Let $expit(u) = \frac{1}{1+exp(-u)}$. An unbiased estimating function of $\theta$, $\phi(\theta)$ , is defined as:
\begin{equation}
\label{ee_syselect}
\begin{pmatrix}
    (\frac{V_i T_i}{e_i \pi_{V,i}} - \frac{V_i(1-T_i)}{(1-e_i)\pi_{V,i}})Y_i - \tau^S (Y_{\mathcal{V}}, \cdot) \\
    (T_i - expit(\gamma X_i))(expit(\eta \tilde{X_i})X_i^T \\
    (V_i - expit(\eta \tilde{X_i}))\tilde{X_i}^T \\
    R_i(Y_i^* - \alpha - (p_{11} - p_{10})\beta T_i) \\
    R_i T_i(Y_i^* - \alpha - (p_{11} - p_{10})\beta T_i ) \\
    (Y_i Y_i^* - p_{11}Y_i) V_i \frac{n}{n_\mathbf{V}} \\
    ( (1-Y_i)Y_i^* - p_{10}(1-Y_i))V_i \frac{n}{n_\mathbf{V}}
    
\end{pmatrix}    
\end{equation}
With this approach, we can obtain sandwich standard errors estimates for $\hat{\tau}^S (Y_{\mathcal{V}}, \cdot)$ and $\hat{\tau}^S (Y_{\mathcal{V}}, Y^*_{\mathcal{V}^C})$.

We can define a similar set of estimating equations for $\hat{\tau}^S(Y_{\mathcal{V}}, Y^*)$ and $\hat{\tau}^S_{opt}(Y_{\mathcal{V}}, Y^*)$ while considering all $\bm{Y}^*$ for directly estimating the ATE, rather than only a subset. As described in the main paper, both of these estimators are weighted combinations of $\hat{\tau}^S (Y_{\mathcal{V}}, \cdot)$ and $\hat{\tau}(\cdot, Y^*)$. Since estimation of $\hat{\tau}(\cdot, Y^*)$ does not involve accounting for validation sample propensities, we can denote $\gamma_S$ for the parameters used in estimating the accompanying treatment propensity model. Furthermore, define:
\begin{equation}
    D_i = \frac{T_i}{e_i} + \frac{1-T_i}{1-e_i}
\end{equation}

Thus, we consider $\theta =( \tau^S (Y_{\mathcal{V}}, \cdot),\gamma^T, \eta^T ,\alpha, \beta, p_{11}, p_{10}, \gamma_S^T)^T$. An unbiased estimating function of $\theta$ for these estimators, $\phi(\theta)$, is defined as:
\begin{equation}
\label{newSelect}
\begin{pmatrix}
    (\frac{V_i T_i}{e_i \pi_{V,i}} - \frac{V_i(1-T_i)}{(1-e_i)\pi_{V,i}})Y_i - \tau^S (Y_{\mathcal{V}}, \cdot) \\
    (T_i - expit(\gamma X_i))(expit(\eta \tilde{X_i})X_i^T \\
    (V_i - expit(\eta \tilde{X_i}))\tilde{X_i}^T \\
    D_i(Y_i^* - \alpha - (p_{11} - p_{10})\beta T_i) \\
    D_i T_i(Y_i^* - \alpha - (p_{11} - p_{10})\beta T_i ) \\
    (Y_i Y_i^* - p_{11}Y_i) V_i \frac{n}{n_\mathbf{V}} \\
    ( (1-Y_i)Y_i^* - p_{10}(1-Y_i))V_i \frac{n}{n_\mathbf{V}}\\
    (T_i - expit(\gamma_S X_i))X_i^T
\end{pmatrix}
\end{equation}
With this approach, we can obtain sandwich standard errors estimates for $\hat{\tau}^S (Y_{\mathcal{V}}, \cdot)$, $\hat{\tau}(\cdot, Y^*)$, and $\hat{\tau}^S(Y_{\mathcal{V}}, Y^*)$ (or $\hat{\tau}^S_{opt}(Y_{\mathcal{V}}, Y^*)$). 

\FloatBarrier

\newpage

\section{Estimation of Coverage}
As described in the main paper, an estimate of the true ATE was obtained by generating large data sets ($n=50,000$) from the true model, calculating the IPW estimate of the ATE, and taking the average across 5,000 iterations. We can denote this estimate of the truth as $\hat{\tau_0}$. Let $\xi$ represent a given estimator (e.g. $\hat{\tau}^S (Y_{\mathcal{V}}, \cdot)$, $\hat{\tau}(\cdot, Y^*)$, etc.). We can construct a $(1-\alpha)\%$ confidence interval using the sandwich variance estimate of $\xi$, $\hat{V}ar(\xi)$ and the $z$ score from the standard normal distribution as follows: 
\begin{equation}
    \hat{\xi} \pm z_{1-\frac{\alpha}{2}}\sqrt{\hat{V}ar(\xi)}
\end{equation}.

We then assess the coverage of $\hat{\tau_0}$ with the confidence interval for each iteration and average over iterations to obtain the final estimate of coverage.

\section{Additional Simulation Results}

The following sections contain simulation results for a range of additional scenarios: modifying sample sizes for the validation sample; considering more extreme levels of misclassification; varying the degree of validation sample selection bias; and simulating scenarios with heterogeneous misclassification probabilities. These additional simulation results follow from 5000 MC iterations.

\subsection{Varying Validation Sample Sizes}

While maintaining $p_{10}=0.24$, we considered $n_\mathcal{V} \in \{500, 1500\}$. Results are shown in Web Table \ref{tab:sim_varyV_SRS} and Web Table \ref{tab:sim_varyV_np}. For both smaller and greater validation sample sizes that were non-probability samples, our proposed estimators produced unbiased estimates of the ATE with smaller variance.



\begin{table}[htb]
    \caption{Simulation results for bias, average sandwich SE, and 95\% confidence interval coverage probabilities for varying validation sizes $n_{\mathcal{V}}$ while maintaining misclassification probabilities $p_{10} = 0.24$. Results are reported for scenarios in which the validation sample is a SRS.}
    \label{tab:sim_varyV_SRS}
    \centering
    \begin{tabular}{llccc}
    \hline
    \textbf{$n_\mathcal{V}$} & \textbf{Estimator} & \textbf{Bias}  & \textbf{Average } & \textbf{Coverage} \\
    &  &  & \textbf{Sandwich SE} & \\
    \hline
    500 & $\hat{\tau}$ & 0.000 & 0.011  & 0.948 \\
    & $\hat{\tau} (Y_{\mathcal{V}}, \cdot)$ & 0.000 & 0.039  & 0.921 \\
    & $\hat{\tau} (Y_{\mathcal{V}}, Y^*_{\mathcal{V}^C})$ & 0.001 & 0.037 & 0.959 \\
    & $\hat{\tau}^S (Y_{\mathcal{V}}, Y^*_{\mathcal{V}^C})$ & 0.002 & 0.034  & 0.947 \\
    & $\hat{\tau}^S (Y_{\mathcal{V}}, \cdot)$ & 0.000 & 0.039  & 0.922 \\
    & $\hat{\tau}(\cdot, Y^*)$ & 0.001 & 0.036  & 0.946 \\
    &  $\hat{\tau}^S(Y_{\mathcal{V}}, Y^*)$ & 0.001 & 0.033  & 0.943 \\
    & $\hat{\tau}^S_{opt}(Y_{\mathcal{V}}, Y^*)$ & 0.004 & 0.027 & 0.919 \\
    
    \hline
    1500 & $\hat{\tau}$ & 0.000 & 0.011  & 0.948 \\
    & $\hat{\tau} (Y_{\mathcal{V}}, \cdot)$ & 0.000 & 0.022 & 0.943 \\
    & $\hat{\tau} (Y_{\mathcal{V}}, Y^*_{\mathcal{V}^C})$ & 0.000 & 0.032  & 0.951 \\
    & $\hat{\tau}^S (Y_{\mathcal{V}}, Y^*_{\mathcal{V}^C})$ & 0.001 & 0.030  & 0.948 \\
    & $\hat{\tau}^S (Y_{\mathcal{V}}, \cdot)$ & 0.000 & 0.022  & 0.946 \\
    & $\hat{\tau}(\cdot, Y^*)$ & 0.000 & 0.035  & 0.950 \\
    & $\hat{\tau}^S(Y_{\mathcal{V}}, Y^*)$ & 0.000 & 0.026  & 0.946 \\
    & $\hat{\tau}^S_{opt}(Y_{\mathcal{V}}, Y^*)$ & 0.001  & 0.020 & 0.944 \\
    \hline
    \\
    \end{tabular}
\end{table}
\FloatBarrier

\begin{table}[t]
    \caption{Simulation results for bias, average sandwich SE, and 95\% confidence interval coverage probabilities for varying validation sizes $n_{\mathcal{V}}$ while maintaining misclassification probabilities $p_{10} = 0.24$. Results are reported for scenarios in which the validation sample is a non-probability sample.}
    \label{tab:sim_varyV_np}
    \centering
    \begin{tabular}{llccc}
    \hline
    \textbf{$n_\mathcal{V}$} & \textbf{Estimator} & \textbf{Bias}  & \textbf{Average } & \textbf{Coverage} \\
    &  &  & \textbf{Sandwich SE} & \\
    \hline
    500 & $\hat{\tau}$ & 0.000 & 0.011  & 0.948 \\
    & $\hat{\tau}(Y_{\mathcal{V}}, \cdot)$ & 0.155 & 0.072  & 0.429 \\
    & $\hat{\tau}(Y_{\mathcal{V}}, Y^*_{\mathcal{V}^C})$ & -0.017 & 0.034 & 0.924 \\
    & $\hat{\tau}^S (Y_{\mathcal{V}}, Y^*_{\mathcal{V}^C})$ & 0.001 & 0.038 & 0.949 \\
    & $\hat{\tau}^S (Y_{\mathcal{V}}, \cdot)$ & 0.001 & 0.035  & 0.942 \\
    & $\hat{\tau}(\cdot, Y^*)$ & 0.001 & 0.035  & 0.948 \\
    & $\hat{\tau}^S(Y_{\mathcal{V}}, Y^*)$ & 0.001 & 0.032  & 0.950 \\
    & $\hat{\tau}^S_{opt}(Y_{\mathcal{V}}, Y^*)$ & 0.003 & 0.024 & 0.933 \\
    \hline
    1500 & $\hat{\tau}$ & 0.000 & 0.011  & 0.948 \\
    & $\hat{\tau} (Y_{\mathcal{V}}, \cdot)$ & 0.080 & 0.031 & 0.294 \\
    & $\hat{\tau} (Y_{\mathcal{V}}, Y^*_{\mathcal{V}^C})$ & -0.037 & 0.029  & 0.763 \\
    & $\hat{\tau}^S (Y_{\mathcal{V}}, Y^*_{\mathcal{V}^C})$ & 0.002 & 0.045  & 0.933 \\
    & $\hat{\tau}^S (Y_{\mathcal{V}}, \cdot)$ & 0.000 & 0.016  & 0.936 \\
    & $\hat{\tau}(\cdot, Y^*)$ & 0.000 & 0.035  & 0.953 \\
    & $\hat{\tau}^S(Y_{\mathcal{V}}, Y^*)$ & 0.000 & 0.025  & 0.950 \\
    & $\hat{\tau}^S_{opt}(Y_{\mathcal{V}}, Y^*)$ & 0.000  & 0.015 & 0.931 \\
    \hline
    \\
    \end{tabular}
\end{table}
\FloatBarrier

\subsection{Varying Misclassification Rates}

In the main paper, we considered $p_{10} = 0.24$ for $n_\mathcal{V} \approx 850$. Having a misclassification probability of $p_{10} = 0.24$ already represents a scenario with a fairly high degree of misclassification. To explore even more extreme levels of misclassification while maintaining  $n_\mathcal{V} \approx 850$, we considered $p_{10} \in \{0.16, 0.32 \}$. Results are shown in Web Table \ref{tab:sim_varyME_SRS} and Web Table \ref{tab:sim_varyME_np}. Especially in the case when the validation samples are non-probability samples, we observe that our proposed estimators are robust to more extreme levels of misclassification.

\begin{table}[htb]
    \caption{Simulation results for bias, average sandwich SE, and 95\% confidence interval coverage probabilities for varying misclassification probabilities $p_{10}$ while maintaining $n_\mathcal{V} = 850$. Results are reported for scenarios in which the validation sample is a SRS.}
    \label{tab:sim_varyME_SRS}
    \centering
    \begin{tabular}{llccc}
    \hline
    \textbf{$p_{10}$} & \textbf{Estimator} & \textbf{Bias}  & \textbf{Average} & \textbf{Coverage} \\
     & & & \textbf{Sandwich SE} &  \\
    \hline
    {0.16} & $\hat{\tau}$ & 0.000 & 0.011  & 0.948 \\
    &  {$\hat{\tau} (Y_{\mathcal{V}}, \cdot)$} & 0.000 & 0.030 & 0.935 \\
    & {$\hat{\tau} (Y_{\mathcal{V}}, Y^*_{\mathcal{V}^C})$} & 0.000 & 0.027 & 0.956 \\
    & {$\hat{\tau}^S (Y_{\mathcal{V}}, Y^*_{\mathcal{V}^C})$} & 0.000 & 0.026 & 0.949 \\
    & {$\hat{\tau}^S (Y_{\mathcal{V}}, \cdot)$} & 0.000 & 0.030 & 0.941 \\
    & {$\hat{\tau}(\cdot, Y^*)$} & 0.000 & 0.027 & 0.949 \\
    & {$\hat{\tau}^S(Y_{\mathcal{V}}, Y^*)$} & 0.000 & 0.024 & 0.950 \\
    & {$\hat{\tau}^S_{opt}(Y_{\mathcal{V}}, Y^*)$} & 0.002 & 0.021 & 0.939 \\
    \hline
    {0.32} & $\hat{\tau}$ & 0.000 & 0.011  & 0.948 \\
    &  {$\hat{\tau} (Y_{\mathcal{V}}, \cdot)$} & 0.000 & 0.030 & 0.935 \\
    & {$\hat{\tau} (Y_{\mathcal{V}}, Y^*_{\mathcal{V}^C})$} & 0.002 & 0.046 & 0.957 \\
    & {$\hat{\tau}^S (Y_{\mathcal{V}}, Y^*_{\mathcal{V}^C})$} & 0.002 & 0.042 & 0.948 \\
    & {$\hat{\tau}^S (Y_{\mathcal{V}}, \cdot)$} & 0.000 & 0.030 & 0.941 \\
    & {$\hat{\tau}(\cdot, Y^*)$} & 0.001 & 0.046 & 0.950 \\
    & {$\hat{\tau}^S(Y_{\mathcal{V}}, Y^*)$} & 0.001 & 0.039 & 0.948 \\
    & {$\hat{\tau}^S_{opt}(Y_{\mathcal{V}}, Y^*)$} & 0.001 & 0.026 & 0.935 \\ 
    
    \hline
    \\
    \end{tabular}
\end{table}

\FloatBarrier

\begin{table}[t]
    \caption{Simulation results for bias, average sandwich SE, and 95\% confidence interval coverage probabilities for varying misclassification probabilities $p_{10}$ while maintaining $n_\mathcal{V} \approx 850$. Results are reported for scenarios in which the validation sample is a non-probability sample.}
    \label{tab:sim_varyME_np}
    \centering
    \begin{tabular}{llccc}
    \hline
    \textbf{$p_{10}$} & \textbf{Estimator} & \textbf{Bias}  & \textbf{Average} & \textbf{Coverage} \\
     & & & \textbf{Sandwich SE} &  \\
    \hline
    {0.16} & $\hat{\tau}$ & 0.000 & 0.011  & 0.948 \\
    &  {$\hat{\tau} (Y_{\mathcal{V}}, \cdot)$} & 0.119 & 0.048 & 0.318 \\
    & {$\hat{\tau} (Y_{\mathcal{V}}, Y^*_{\mathcal{V}^C})$} & -0.015 & 0.025 & 0.909 \\
    & {$\hat{\tau}^S (Y_{\mathcal{V}}, Y^*_{\mathcal{V}^C})$} & 0.000 & 0.033 & 0.947 \\
    & {$\hat{\tau}^S (Y_{\mathcal{V}}, \cdot)$} & 0.000 & 0.024 & 0.937 \\
    & {$\hat{\tau}(\cdot, Y^*)$} & 0.000 & 0.027 & 0.951 \\
    & {$\hat{\tau}^S(Y_{\mathcal{V}}, Y^*)$} & 0.000 & 0.023 & 0.948 \\
    & {$\hat{\tau}^S_{opt}(Y_{\mathcal{V}}, Y^*)$} & 0.001 & 0.018 & 0.935 \\
    \hline
    {0.32} & $\hat{\tau}$ & 0.000 & 0.011  & 0.948 \\
    &  {$\hat{\tau} (Y_{\mathcal{V}}, \cdot)$} & 0.119 & 0.048 & 0.318 \\
    & {$\hat{\tau} (Y_{\mathcal{V}}, Y^*_{\mathcal{V}^C})$} & -0.043 & 0.042 & 0.825 \\
    & {$\hat{\tau}^S (Y_{\mathcal{V}}, Y^*_{\mathcal{V}^C})$} & 0.001 & 0.051 & 0.943 \\
    & {$\hat{\tau}^S (Y_{\mathcal{V}}, \cdot)$} & 0.000 & 0.024 & 0.937\\
    & {$\hat{\tau}(\cdot, Y^*)$} & 0.001 & 0.046 & 0.951 \\
    & {$\hat{\tau}^S(Y_{\mathcal{V}}, Y^*)$} & 0.001 & 0.038 & 0.949 \\
    & {$\hat{\tau}^S_{opt}(Y_{\mathcal{V}}, Y^*)$} & 0.001 & 0.021 & 0.934\\
    \hline
    \\
    \end{tabular}
\end{table}

\FloatBarrier
\subsection{Varying Validation Sample Selection Bias}

In the main paper, we simulated a non-probability validation sample by defining $\widetilde{X}_{i} = (1, T_{i}, \mathbf{X}_{i}^{T})^T$ and $\alpha_{0} = (\alpha_{intercept}, 0.5, 1, 1, 1, 1, 0)$. Then $\pi_V(\mathbf{X}_{i}) = expit(\alpha_{0}^T\widetilde{X}_{i})$, where the choice of $\alpha_{intercept}$ varies to achieve targeted values for $n_{\mathcal{V}}$. Here, we consider defining $\alpha_{0}$ in two alternate ways while continuing to target $n_{\mathcal{V}} = 850$ and using misclassification probabilities of $p_{11} = 0.67$ and $p_{10} = 0.24$. 

In the first alternative case, we flip the sign of the coefficients in the original model and consider $\alpha_{0} = (-2.2, -0.5, -1, -1, -1, -1, 0)$. We were motivated to explore (1) to see whether bias resulting from measurement error in the outcome could be counteracting validation sample selection bias. If these two sources of bias counteracted each other, approaches that do not account for sample selection bias could appear to have smaller magnitudes of bias than actuality.In the second alternative case, we increase the magnitude of coefficients and consider $\alpha_{0} = (-4, 1, 1.5, 1.5, 1.5, 1.5, 0)$.  The results are shown in Web Table \ref{tab:sim_varyaltV_np}. Flipping the signs of the coefficient in our validation sample selection model seems to confirm that biases are not unintentionally mitigated under approaches that do not account for sample selection bias. The estimates of bias were minimized by our proposed estimators, as we might expect. When the magnitude of the coefficients are increased for the validation sample selection model, the estimators that do not correct for sample selection bias yield a greater magnitude of bias in estimates of the ATE.

\begin{table}[!htb]
    \caption{Simulation results for bias, average sandwich SE, and 95\% confidence interval coverage probabilities for varying models for the validation non-probability sample selection. In generated data sets, we maintained a misclassification probability of $p_{10} = 0.24$ and $n_\mathcal{V} \approx 850$.}
    \label{tab:sim_varyaltV_np}
    \centering
    \begin{tabular}{llccc}
    \hline
    \textbf{$\alpha_{0}$} & \textbf{Estimator} & \textbf{Bias}  & \textbf{Average} & \textbf{Coverage} \\
    & &  & \textbf{Sandwich SE} & \\
    \hline
    {$(-2.2, -0.5, -1, -1, -1, -1, 0)$} & $\hat{\tau}$ & 0.000 & 0.011  & 0.948 \\
    &  {$\hat{\tau} (Y_{\mathcal{V}}, \cdot)$} & -0.060 & 0.010 & 0.000 \\
    & {$\hat{\tau} (Y_{\mathcal{V}}, Y^*_{\mathcal{V}^C})$} & 0.039 & 0.085 & 0.986 \\
    & {$\hat{\tau}^S (Y_{\mathcal{V}}, Y^*_{\mathcal{V}^C})$} & 0.006 & 0.052 & 0.943 \\
    & {$\hat{\tau}^S (Y_{\mathcal{V}}, \cdot)$} & -0.004 & 0.126 & 0.960 \\
    & {$\hat{\tau}(\cdot, Y^*)$} & 0.009 & 0.041 & 0.921 \\
    & {$\hat{\tau}^S(Y_{\mathcal{V}}, Y^*)$} & 0.006 & 0.045 & 0.931 \\
    & {$\hat{\tau}^S_{opt}(Y_{\mathcal{V}}, Y^*)$} & 0.004 & 0.033 & 0.911\\
    \hline
    {$(-4, 1, 1.5, 1.5, 1.5, 1.5, 0)$} & $\hat{\tau}$ & 0.000 & 0.011  & 0.948 \\
    &  {$\hat{\tau} (Y_{\mathcal{V}}, \cdot)$} & 0.160 & 0.050 & 0.149 \\
    & {$\hat{\tau} (Y_{\mathcal{V}}, Y^*_{\mathcal{V}^C})$} & -0.038 & 0.032 & 0.771 \\
    & {$\hat{\tau}^S (Y_{\mathcal{V}}, Y^*_{\mathcal{V}^C})$} & 0.002 & 0.050 & 0.948\\
    & {$\hat{\tau}^S (Y_{\mathcal{V}}, \cdot)$} & 0.000 & 0.031 & 0.938\\
    & {$\hat{\tau}(\cdot, Y^*)$} & 0.000 & 0.035 & 0.950 \\
    & {$\hat{\tau}^S(Y_{\mathcal{V}}, Y^*)$} & 0.000 & 0.030 & 0.951 \\
    & {$\hat{\tau}^S_{opt}(Y_{\mathcal{V}}, Y^*)$} & 0.002 & 0.021 & 0.926\\
    \hline
    \\
    \end{tabular}
\end{table}

\FloatBarrier
\subsection{Assuming Heterogeneous Misclassification Probabilities} 

Another consideration with the estimators we proposed in the main paper is the assumption of homogeneous misclassification probabilities, which may not hold with real world data. To explore the performance of estimators with heterogeneous misclassification probabilities, we simulated different misclassification probabilities for the two treatment groups. For simulating heterogeneous misclassification probabilities, consider ${p_{10}}(T) = expit(-2 + 0.5T)$ such that $(Y^*|Y=0,T=t) \sim Bern(expit(-2+.5t))$ for $t\in {0,1}$. Results are shown in Web Table \ref{tab:sim_het2_res}. We have preliminary evidence that our estimators display some robustness to heterogenous misclassification probabilities, where our proposed estimators continue to lead to unbiased estimates of the ATE with close to nominal coverage. However, we anticipate that our estimators would exhibit bias under more extreme violations of the homogeneous misclassification assumption. Future work is needed to more extensively evaluate the degree to which violating the homogeneous misclassification probability assumption impacts estimation of the ATE with these estimators. It is probable that our estimators would exhibit bias in settings with more extreme violations of the homogeneous misclassification assumption.


\begin{table}[!htb]
    \caption{Simulation results for bias, average sandwich SE, and 95\% confidence interval coverage probabilities for varying models for the validation non-probability sample selection. In generated data sets, we maintained a misclassification probability of $p_{10} = 0.24$ and $n_\mathcal{V} \approx 850$.}
    \label{tab:sim_het2_res}
    \centering
    \begin{tabular}{llccc}
    \hline
    \textbf{Validation} & \textbf{Estimator} & \textbf{Bias}  & \textbf{Average} & \textbf{Coverage} \\
    \textbf{Sample Type} & & & \textbf{Sandwich SE} & \\
    \hline
    {SRS} & $\hat{\tau}$ & 0.000 & 0.011  & 0.948 \\
    &  {$\hat{\tau} (Y_{\mathcal{V}}, \cdot)$} & 0.000 & 0.030 & 0.935 \\
    & {$\hat{\tau} (Y_{\mathcal{V}}, Y^*_{\mathcal{V}^C})$} & 0.000 & 0.027 & 0.952 \\
    & {$\hat{\tau}^S (Y_{\mathcal{V}}, Y^*_{\mathcal{V}^C})$} & 0.000 & 0.026 & 0.947 \\
    & {$\hat{\tau}^S (Y_{\mathcal{V}}, \cdot)$} & 0.000 & 0.030 & 0.941 \\
    & {$\hat{\tau}(\cdot, Y^*)$} & 0.000 & 0.028 & 0.949 \\
    & {$\hat{\tau}^S(Y_{\mathcal{V}}, Y^*)$} & 0.000 & 0.024 & 0.948 \\
    & {$\hat{\tau}^S_{opt}(Y_{\mathcal{V}}, Y^*)$} & 0.002 & 0.021 & 0.938\\
    \hline
    {Non-probability} & $\hat{\tau}$ & 0.000 & 0.011  & 0.948 \\
    &  {$\hat{\tau} (Y_{\mathcal{V}}, \cdot)$} & 0.119 & 0.048 & 0.318 \\
    & {$\hat{\tau} (Y_{\mathcal{V}}, Y^*_{\mathcal{V}^C})$} & -0.015 & 0.025 & 0.907 \\
    & {$\hat{\tau}^S (Y_{\mathcal{V}}, Y^*_{\mathcal{V}^C})$} &  0.000 & 0.033 & 0.948\\
    & {$\hat{\tau}^S (Y_{\mathcal{V}}, \cdot)$} & 0.000 & 0.024 & 0.937\\
    & {$\hat{\tau}(\cdot, Y^*)$} & 0.000 & 0.027 & 0.951 \\
    & {$\hat{\tau}^S(Y_{\mathcal{V}}, Y^*)$} & 0.000 & 0.023 & 0.946 \\
    & {$\hat{\tau}^S_{opt}(Y_{\mathcal{V}}, Y^*)$} & 0.001 & 0.018 & 0.934\\
    \hline
    \\
    \end{tabular}
\end{table}
\FloatBarrier
\subsection{Misspecification of Validation Sample Selection Model} 

While we assume that the validation sample selection mechanism is correctly specified, potential misspecification of the model could exist with real world analyses. To explore how our estimators perform with a misspecified selection model, we conducted additional simulations where our validation sample selection model specification excluded one of the pertinent covariates, $X_2$. Results are shown in Web Table \ref{tab:sim_missV_np}. As we might expect, estimates under all the estimators are biased, especially estimators that rely solely on the validation sample. All estimators exhibited under-coverage, although one of our proposed estimators, $\hat{\tau}^S (Y_{\mathcal{V}}, Y^*_{\mathcal{V}^C})$, had the least amount of bias and coverage closest to nominal.

\begin{table}[!htb]
    \caption{Simulation results for bias, average sandwich SE, and 95\% confidence interval coverage probabilities with a misspecified model for selection into the validation non-probability sample.}
    \label{tab:sim_missV_np}
    \centering
    \begin{tabular}{llccc}
    \hline
    \textbf{Validation} & \textbf{Estimator} & \textbf{Bias}  & \textbf{Average} & \textbf{Coverage} \\
    \textbf{Sample Type} & & & \textbf{Sandwich SE} & \\
    \hline
    Excluding  & $\hat{\tau}$ & 0.000 & 0.011  & 0.948 \\
    $X_2$ &  {$\hat{\tau} (Y_{\mathcal{V}}, \cdot)$} & 0.178 & 0.043 & 0.033 \\
    (Misspecified) & {$\hat{\tau} (Y_{\mathcal{V}}, Y^*_{\mathcal{V}^C})$} & -0.020 & 0.032 & 0.903 \\
    & {$\hat{\tau}^S (Y_{\mathcal{V}}, Y^*_{\mathcal{V}^C})$} & 0.015 & 0.035 & 0.920 \\
    & {$\hat{\tau}^S (Y_{\mathcal{V}}, \cdot)$} & 0.038 & 0.036 & 0.740 \\
    & {$\hat{\tau}(\cdot, Y^*)$} & 0.021 & 0.035 & 0.911 \\
    & {$\hat{\tau}^S(Y_{\mathcal{V}}, Y^*)$} & 0.024 & 0.030 & 0.873 \\
    & {$\hat{\tau}^S_{opt}(Y_{\mathcal{V}}, Y^*)$} & 0.031 & 0.026 & 0.731\\
    \hline
    \\
    \end{tabular}
\end{table}

\FloatBarrier
\section{Additional Details for Real Data Analysis}
For the real data analysis with ACT, coefficient estimates for the propensity score models for hypertension and autopsy sample selection are presented in Web Table \ref{tab:act_modT} and Web Table \ref{tab:act_modV}, respectively. Standardized mean differences (SMD) based on hypertension status also were obtained for the entire ACT cohort as well (Web Table \ref{tab:act_hypertension_smdAll}) as the autopsy sub-sample (Web Table \ref{tab:act_hypertension_smdAutopsy}). Furthermore, a summary of characteristics following stratification by autopsy status for the entire ACT cohort is shown in Web Table \ref{tab:act_smdAutopsy}.

\begin{table}[!htb]
    \caption{Coefficient estimates with 95\% confidence intervals on the log-odds scale for the hypertension propensity score model using ACT data.}
    \label{tab:act_modT}
    \centering
    \begin{tabularx}{\textwidth}{X*{2}{>{\centering\arraybackslash}X}}
    \hline
    \textbf{Variable} & \textbf{Coefficient} & \textbf{95\% CI}\\
    \hline
    Intercept & -1.345 & (-1.767, -0.927)\\
    \hline
    Previous or Current Hypertension Medication Use  & 0.935 & (0.81, 1.06)\\
    \hline
    BMI & 0.033 & (0.021, 0.046)\\
    \hline
    Age Group at Last Visit (years) & & \\
    (74, 80] & 0.528 & (0.361, 0.697)\\
    
    (80, 84] & 0.935 & (0.744, 1.127)\\
    
    (84, 89] & 1.205 & (1.02, 1.392)\\
    
    $>$89 & 1.421 & (1.216, 1.631)\\
    \hline
    White & -0.244 & (-0.45, -0.041)\\
    \hline
    Female & 0.221 & (0.1, 0.343)\\
    \hline
    \\
    \end{tabularx}
\end{table}

\begin{table}[!htb]
    \caption{Coefficient estimates with 95\% confidence intervals on the log-odds scale for the autopsy sample selection propensity score model using ACT data.}
    \label{tab:act_modV}
    \centering
    \begin{tabularx}{\textwidth}{X*{2}{>{\centering\arraybackslash}X}}
    \hline
    \textbf{Variable} & \textbf{Coefficient} & \textbf{95\% CI}\\
    \hline
    Intercept & -3.216 & (-3.687, -2.768)\\
    \hline
    Hypertension  & -0.017 & (-0.198, 0.167)\\
    \hline
    Age Group at Last Visit (years) & & \\
    (74, 80] & 0.615 & (0.262, 0.983)\\
    
    (80, 84] & 0.984 & (0.631, 1.354)\\
    
    (84, 89] & 1.085 & (0.744, 1.445)\\
    
    $>$89 & 1.484 & (1.142, 1.846)\\
    \hline
    White & 0.640 & (0.336, 0.965)\\
    \hline
    Female & -0.188 & (-0.346, -0.029)\\
    \hline
    Any Dementia Diagnosis  & 0.850 & (0.688, 1.012)\\
    \hline
    ACT Cohort & & \\
    Expansion & 0.219 & (0.02, 0.416)\\
    
    Replacement & -0.866 & (-1.084, -0.653)\\
    \hline
    \\
    \end{tabularx}
\end{table}

\begin{table}[!htb]
    \caption{Summary of select characteristics, stratifying by hypertension status, for the full ACT cohort. Standardized mean differences (SMD) also are provided.}
    \label{tab:act_hypertension_smdAll}
    \centering
    \begin{tabularx}{\textwidth}{X*{3}{>{\centering\arraybackslash}X}}
    \hline
    \textbf{Characteristic} & \textbf{No Hypertension} & \textbf{Hypertension} & \textbf{SMD} \\
    \hline
    n & 1756 & 3913 & \\
    \hline
    Possible/Probable clinical AD diagnosis (\%) & 247 (14.1) & 855 ( 21.9) & 0.204\\
    \hline
    Previous or Current Hypertension Medication Use (\%) & 814 (46.4) & 2871 ( 73.4) & 0.573\\
    \hline
    Age Group at Last Visit (years) (\%) &  &  & 0.569\\
    
    [65,74] & 577 (32.9) & 554 ( 14.2) & \\
    
    (74,80] & 481 (27.4) & 851 ( 21.7) & \\
    
    (80,84] & 257 (14.6) & 694 ( 17.7) & \\
    
    (84,89] & 265 (15.1) & 970 ( 24.8) & \\
    
    $>$89 & 176 (10.0) & 844 ( 21.6) & \\
    
    \hline
    White (\%) & 1600 (91.1) & 3478 ( 88.9) & 0.074\\
    \hline
    Female (\%) & 936 (53.3) & 2340 ( 59.8) & 0.131\\
    \hline
    Any Dementia Diagnosis  (\%) & 297 (16.9) & 1036 ( 26.5) & 0.234\\
    \hline
    ACT Cohort (\%) &  &  & 0.391\\
    
    Original & 584 (33.3) & 1983 ( 50.7) & \\
    
    Expansion & 234 (13.3) & 551 ( 14.1) & \\
    
    Replacement & 938 (53.4) & 1379 ( 35.2) & \\
    \hline
    BMI (mean (SD)) & 27.40 (5.05) & 28.55 (5.30) & 0.223\\
    \hline
    \\
    \end{tabularx}
\end{table}

\begin{table}[!htb]
    \caption{Summary of select characteristics, stratifying by hypertension status, for the autopsy sub-sample for ACT. Standardized mean differences (SMD) also are provided.}
    \label{tab:act_hypertension_smdAutopsy}
    \centering
    \begin{tabularx}{\textwidth}{X*{3}{>{\centering\arraybackslash}X}}
    \hline
    \textbf{Characteristic} & \textbf{No Hypertension} & \textbf{Hypertension} & \textbf{SMD} \\
    \hline
    n & 205 & 632 & \\
    \hline
    Possible/Probable clinical AD diagnosis (\%) & 74 (36.1) & 243 ( 38.4) & 0.049\\
    \hline
    Previous or Current Hypertension Medication Use (\%) & 102 (49.8) & 483 ( 76.4) & 0.575\\
    \hline
    Age Group at Last Visit (years) (\%) &  &  & 0.290\\
    
    [65,74] & 15 ( 7.3) & 30 (  4.7) & \\
    
    (74,80] & 42 (20.5) & 78 ( 12.3) & \\
    
    (80,84] & 39 (19.0) & 114 ( 18.0) & \\
    
    (84,89] & 56 (27.3) & 188 ( 29.7) & \\
    
    $>$89 & 53 (25.9) & 222 ( 35.1) & \\
    \hline
    White (\%) & 191 (93.2) & 597 ( 94.5) & 0.054\\
    \hline
    Female (\%) & 104 (50.7) & 379 ( 60.0) & 0.187\\
    \hline
    Any Dementia Diagnosis  (\%) & 87 (42.4) & 293 ( 46.4) & 0.079\\
    \hline
    ACT Cohort (\%) &  &  & 0.359\\
    
    Original & 101 (49.3) & 417 ( 66.0) & \\
    
    Expansion & 55 (26.8) & 131 ( 20.7) & \\
    
    Replacement & 49 (23.9) & 84 ( 13.3) & \\
    \hline
    BMI (mean (SD)) & 27.02 (4.91) & 28.02 (4.98) & 0.202\\
    \hline
    \\
    \end{tabularx}
\end{table}

\begin{table}[!htb]
    \caption{Summary of select characteristics, stratifying by status of receiving an autopsy, for the entire cohort of ACT. Standardized mean differences (SMD) also are provided.}
    \label{tab:act_smdAutopsy}
    \centering
    \begin{tabularx}{\textwidth}{X*{3}{>{\centering\arraybackslash}X}}
    \hline
    \textbf{Characteristic} & \textbf{No Autopsy} & \textbf{Autopsy} & \textbf{SMD} \\
    \hline
    n & 4832 & 837 & \\
    \hline
    Possible/Probable clinical AD diagnosis (\%) & 785 (16.2) & 317 (37.9) & 0.502\\
    \hline
    Previous or Current Hypertension Medication Use  (\%) & 3100 (64.2) & 585 (69.9) & 0.122\\
    \hline
    Hypertension (\%) & 3281 (67.9) & 632 (75.5) & 0.169\\
    \hline
    Age Group at Last Visit (years) (\%) &  &  & 0.691\\
    
    [65,74] & 1086 (22.5) & 45 ( 5.4) & \\
    
    (74,80] & 1212 (25.1) & 120 (14.3) & \\
    
    (80,84] & 798 (16.5) & 153 (18.3) & \\
    
    (84,89] & 991 (20.5) & 244 (29.2) & \\
    
    (89,107] & 745 (15.4) & 275 (32.9) & \\
    \hline
    White (\%) & 4290 (88.8) & 788 (94.1) & 0.193\\
    \hline
    Female (\%) & 2793 (57.8) & 483 (57.7) & 0.002\\
    \hline
    Any Dementia Diagnosis  (\%) & 953 (19.7) & 380 (45.4) & 0.570\\
    \hline
    ACT Cohort (\%) &  &  & 0.676\\
    
    Original & 2049 (42.4) & 518 (61.9) & \\
    
    Expansion & 599 (12.4) & 186 (22.2) & \\
    
    Replacement & 2184 (45.2) & 133 (15.9) & \\
    \hline
    BMI (mean (SD)) & 28.27 (5.29) & 27.78 (4.98) & 0.096\\
    \hline
    \\
    \end{tabularx}
\end{table}
\FloatBarrier

\newpage
\bibliographystyle{unsrtnat}
\bibliography{WileyNJD-AMA.bib}